\documentclass{webofc}
\usepackage[varg]{txfonts}

\begin{document}
\title{Measurement of the D-meson nuclear modification factor and\\ elliptic flow in $\PbPb$ collisions at $\pmb{\sqrtsNN=5.02~\TeV}$ with ALICE at the LHC}

\author{\firstname{Fabrizio} \lastname{Grosa}\inst{1,2}\fnsep\thanks{\email{fabrizio.grosa@cern.ch}}
        \firstname{for the ALICE Collaboration} \lastname{}}

\institute{Politecnico di Torino, Corso Duca degli Abruzzi 24, 10129 Torino Italy \and INFN sez. Torino, via Pietro Giuria 1, 10125 Torino Italy}

\abstract{%
Heavy-flavour hadrons 
are recognised as a powerful probe for the characterisation of the deconfined medium created in heavy-ion collisions, the Quark-Gluon Plasma (QGP). 
The ALICE Collaboration measured the production of $\Dzero$, $\Dplus$, $\Dstar$ and $\Ds$ mesons in $\PbPb$ collisions at $\sqrtsNN=5.02~\TeV$. The measurement of the nuclear modification factor ($\Raa$) provides a strong evidence of the in-medium parton energy loss. 
The comparison between the $\Ds$ and the non-strange D-meson $\Raa$ can help to study the hadronisation mechanism of the charm quark in the QGP. In mid-central collisions, the measurement of the D-meson elliptic flow $v_2$ at low transverse momentum ($\pt$) gives insight into the participation of the charm quark into the collective motion of the system, while at high $\pt$ it constrains the path-length dependence of the energy loss. The $\Ds$ $v_2$, measured for the first time at the LHC, is found to be compatible to that of non-strange D mesons and positive with a significance of about $2.6~\sigma$. The coupling of the charm quark to the light quarks in the underlying medium is further investigated for the first time with the application of the Event-Shape Engineering (ESE) technique to D-meson elliptic flow.   
}
\maketitle
\section{Introduction}
\label{sec:intro}
Heavy quarks (i.e. charm and beauty) are excellent probes for the characterisation of the deconfined medium created in ultra-relativistic heavy-ion collisions, the Quark-Gluon Plasma (QGP). Because of their large mass, heavy quarks are predominantly produced in hard-scattering processes, before the formation of the QGP \cite{Liu:2012ax}. Therefore, they experience the whole system evolution, interacting with the medium constituents via collisional and radiative processes \cite{Braaten:1991we,Gyulassy:1990ye,Baier:1996sk,Zhang:2003wk,Dokshitzer:2001zm}. The observation of a large suppression of the heavy-flavour hadron yields at intermediate/high $\pt$ in central Pb--Pb collisions with respect to those in pp collisions indicated by the measurement of nuclear modification factor $\Raa(\pt)=({\rm d} N_{\rm AA}/{\rm d}\pt)/(\langle\Taa\rangle \cdot {\rm d}\sigma_{\rm pp}/{\rm d}\pt)$ significantly smaller than unity, provides a strong evidence of the in-medium parton energy loss \cite{Adam:2015sza}. The comparison between heavy-flavour and light-flavour 
hadrons gives insight into the colour-charge and quark-mass dependence of the energy loss \cite{Adam:2015nna}. Moreover, it is predicted that a fraction of heavy quarks could hadronise via coalescence in the medium and, therefore, could be sensitive to the enhanced production of strange quarks in high-energy heavy-ion collisions \cite{Kuznetsova:2006bh}. In this scenario, the measurement of heavy-flavour hadrons with strange-quark content (e.g. the $\Ds$ meson) is crucial to understand the modification of the charm-quark hadronisation in the deconfined medium \cite{Adam:2015jda}. 
	Complementary information on the interaction of heavy quarks with the QGP is provided by the measurement of the azimuthal anisotropy in the momentum distribution of heavy-flavour hadrons and, in particular, by the elliptic flow $v_2=\langle \cos(2(\varphi-\Psi_2))\rangle$, which is defined as the second-order harmonic coefficient of the Fourier expansion of the azimuthal distribution with respect to the reaction plane angle $\Psi_2$. The measurement of the $v_2$ at low $\pt$ helps to quantify to which extent heavy quarks are influenced by the collective dynamics of the underlying medium, while at high $\pt$ has the potential to constrain the path-length dependence of the parton energy loss in the QGP \cite{Batsouli:2002qf,Gyulassy:2000gk,Abelev:2014ipa}. 
\section{D-meson reconstruction}
\label{sec:ALICEandDmesons}
\begin{figure}[b]
\centering
\includegraphics[width=5.5cm,clip]{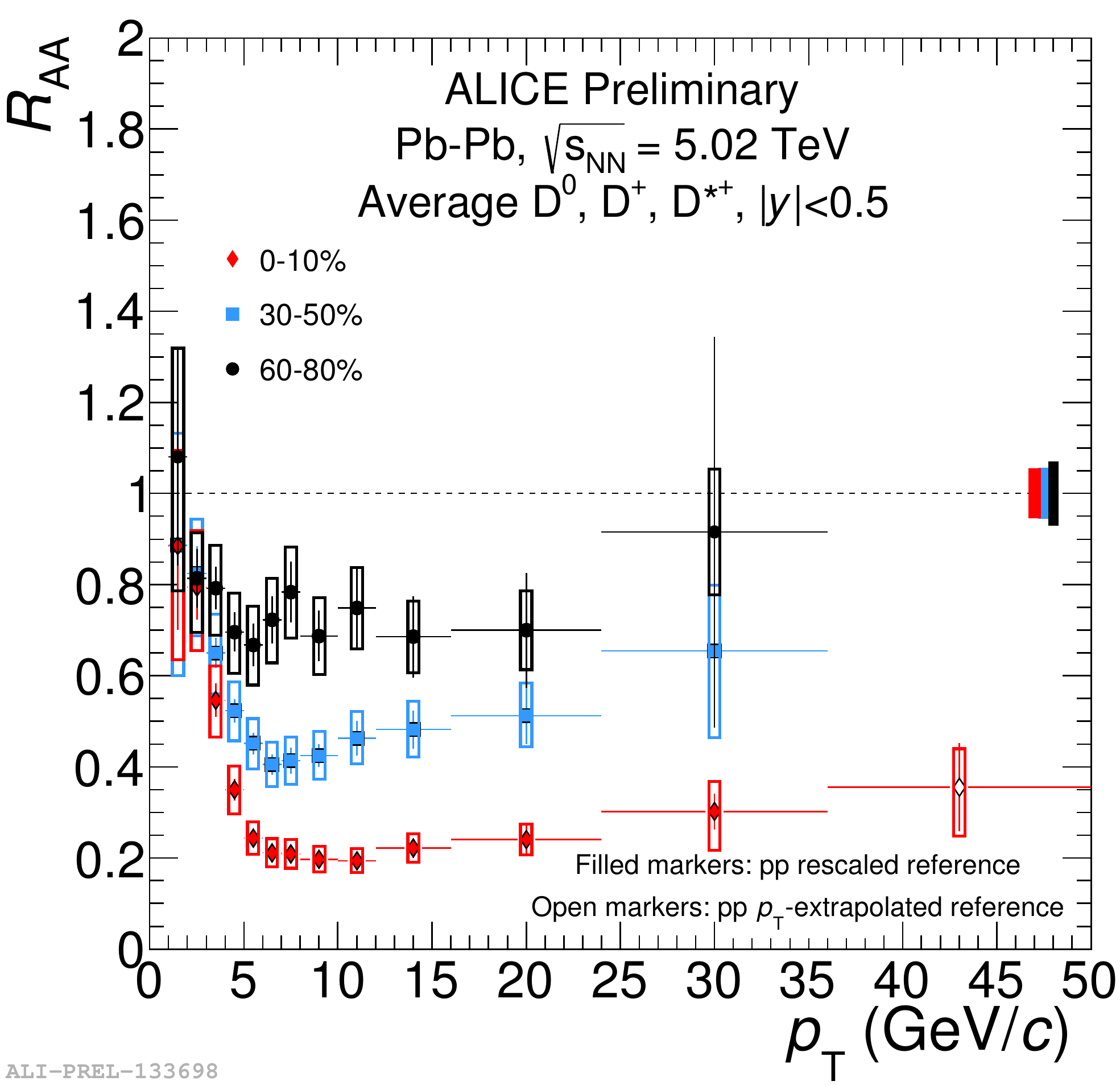}
\hspace{1cm}
\includegraphics[width=5.5cm,clip]{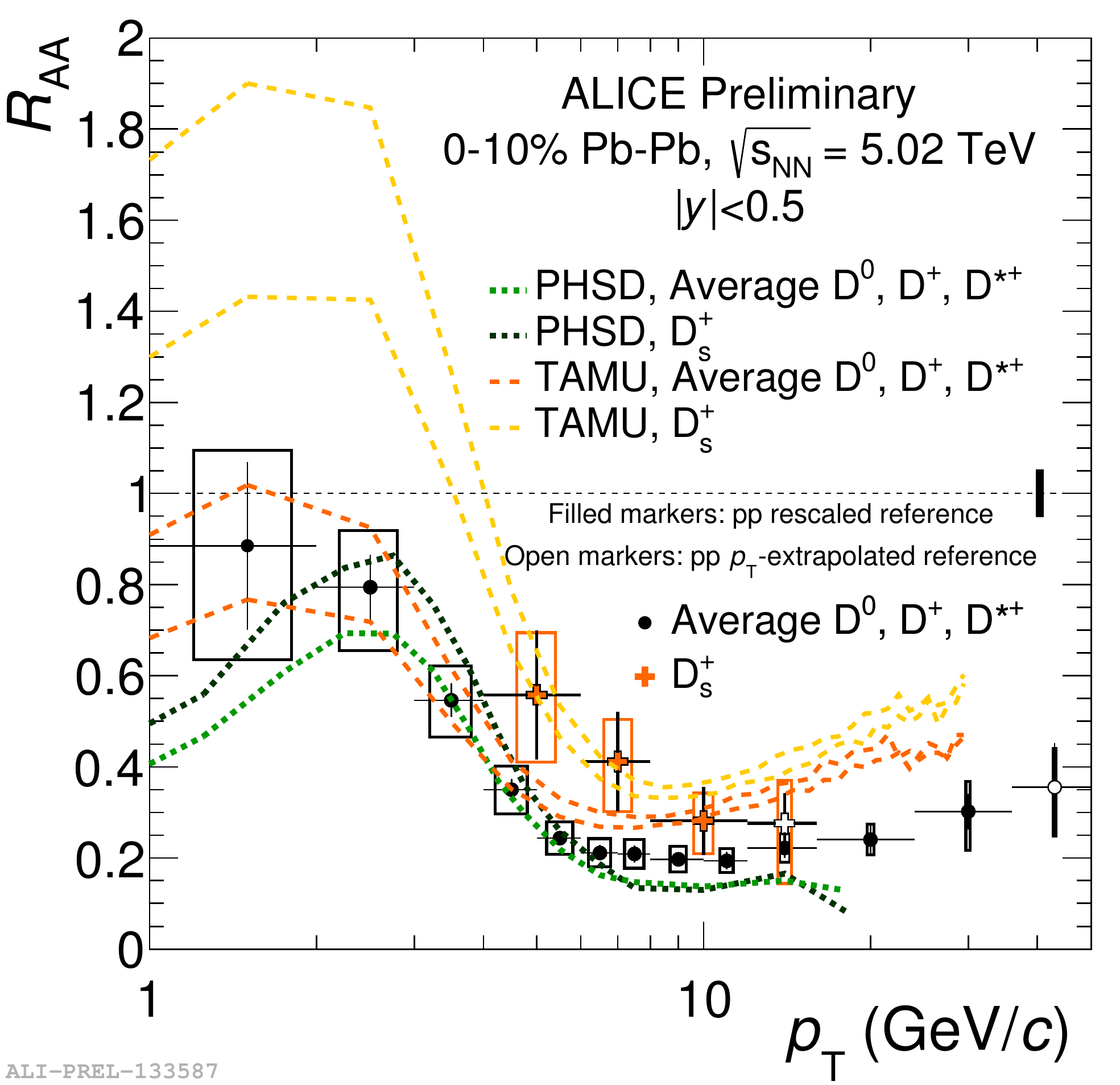}
\caption{Left: average prompt $\Dzero$, $\Dplus$, $\Dstar$ $\pt$-differential $\Raa$ in central 0--10\% (diamonds), mid-central 30--50\% (squares) and peripheral 60--80\% (circles) Pb--Pb collisions at $\sqrtsNN=5.02~\TeV$ \cite{ALICE-PUBLIC-2017-003}. Right: comparison between prompt $\Ds$ (crosses) and non-strange D-meson (circles) $\Raa$ in Pb--Pb collisions at $\sqrtsNN=5.02~\TeV$ in the 0--10\% centrality class \cite{ALICE-PUBLIC-2017-003}. The data are compared to models that include charm hadronisation via coalescence in the QGP \cite{He:2014cla,Song:2015sfa}.}
\label{fig:DmesonRaaAverage}      
\end{figure}

Open-charm production in Pb--Pb collisions at $\sqrtsNN=5.02~\TeV$ was measured by ALICE via the exclusive reconstruction of D mesons at mid-rapidity ($|y|<0.8$), in the hadronic decay channels $\DtoKpi$ ($c\tau\simeq123~\mum$, ${\rm BR}=3.93\%$), $\DtoKpipi$ ($c\tau\simeq312~\mum$, ${\rm BR}=9.46\%$), $\DstartoDpi$ (strong decay, ${\rm BR}=67.7\%$) and $\DstophipitoKKpi$ ($c\tau\simeq150~\mum$, ${\rm BR}=2.67\%$) \cite{Patrignani:2016xqp}. The decay topologies were reconstructed exploiting the excellent vertex-reconstruction capabilities of the Inner Tracking System (ITS). Kaons and pions were identified with the Time Projection Chamber (TPC) via their specific energy loss and with the Time-Of-Flight detector (TOF). The raw D-meson yields were extracted via an invariant-mass analysis after having applied topological selections to enhance the signal over background ratio. The efficiency times acceptance corrections were obtained from MC simulations based on HIJING \cite{Wang:1991hta} and PYTHIA 6 \cite{Sjostrand:2006za} event generators. The fraction of prompt D mesons was estimated with a FONLL-based approach \cite{Cacciari:1998it,Acharya:2017qps}. The centrality and the Event-Plane angle (estimator of $\Psi_2$) were provided by the V0 scintillators, which cover the pseudorapidity regions $-3.7 < \eta < -1.7$ and $2.8 < \eta < 5.1$. 
\section{Prompt D-meson nuclear modification factor and elliptic flow}
\label{sec:RaaAndV2}
The $\pt$-differential $\Raa$ of prompt $\Dzero$, $\Dplus$, $\Dstar$ and $\Ds$ was measured in central 0--10\%, mid-central 30--50\% and peripheral 60--80\% Pb--Pb collisions at $\sqrtsNN=5.02~\TeV$ \cite{ALICE-PUBLIC-2017-003}. The proton-proton reference was obtained  from the measured production cross section at $\sqrts=7~\TeV$ \cite{Acharya:2017jgo}, scaled to $\sqrts=5.02~\TeV$ using the FONLL prediction. The average non-strange D-meson $\Raa$ shows an increasing suppression from peripheral to central events up to a factor about 5 for $\pt>5~\GeV/c$, as reported in the left panel of Figure~\ref{fig:DmesonRaaAverage}. 
The prompt $\Ds$ nuclear modification factor is systematically higher, although compatible within uncertainties, than that of non-strange D mesons in the full $\pt$ range measured, for the three centrality classes. In the right panel of Figure~\ref{fig:DmesonRaaAverage} the $\Ds$ and the non-strange D-meson $\Raa$ in the 0--10\% centrality class are compared to models that include the hadronisation of charm quark via coalescence in a strangeness-enhanced QGP \cite{He:2014cla,Song:2015sfa}. These models predict a smaller suppression of the $\Ds$ with respect to the non-strange D mesons.  
\begin{figure}[t]
\centering
\vspace{-0.78cm}
\includegraphics[width=7.15cm,clip]{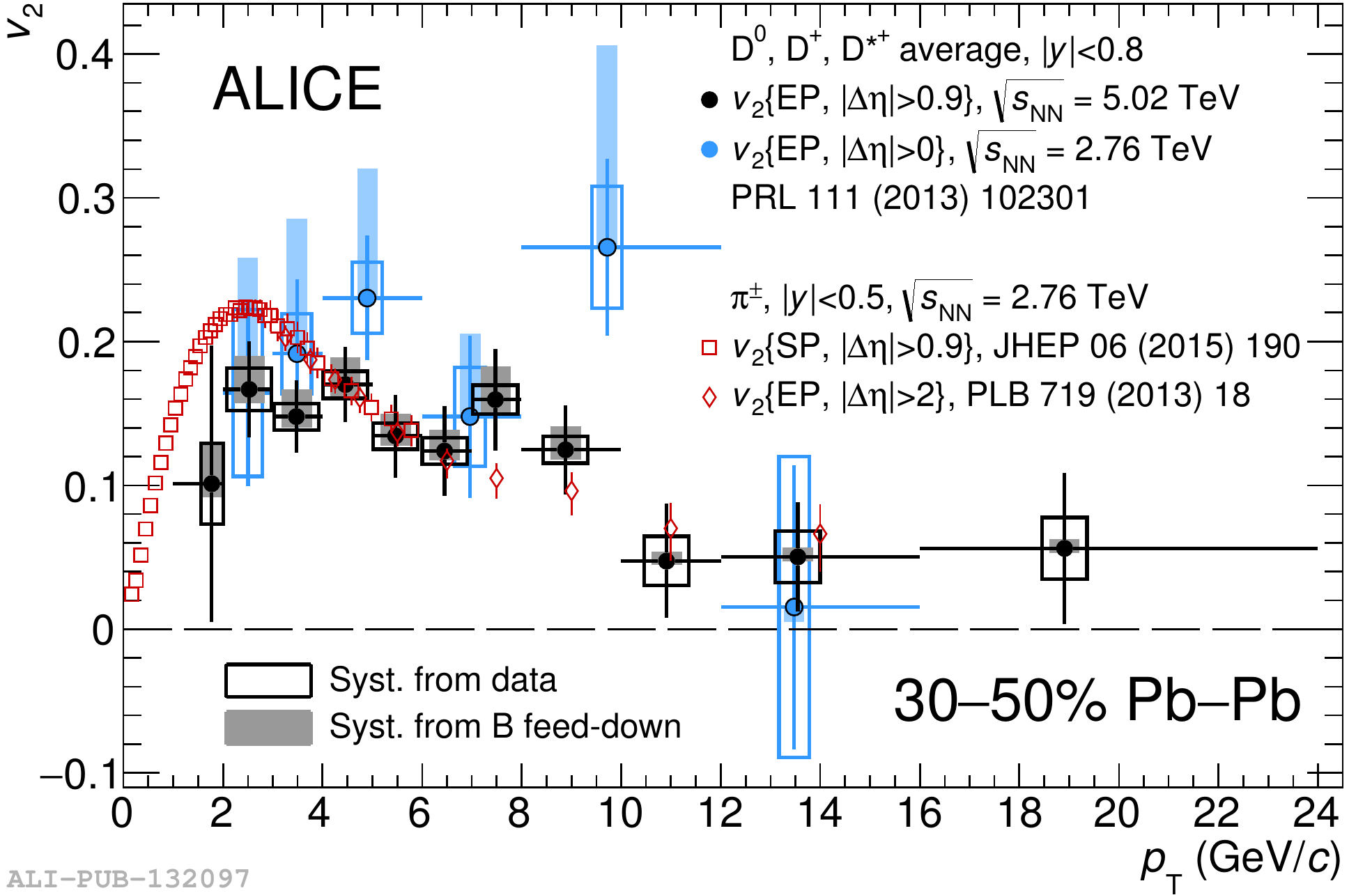}
\hspace{0.5cm}
\includegraphics[width=5.7cm,clip]{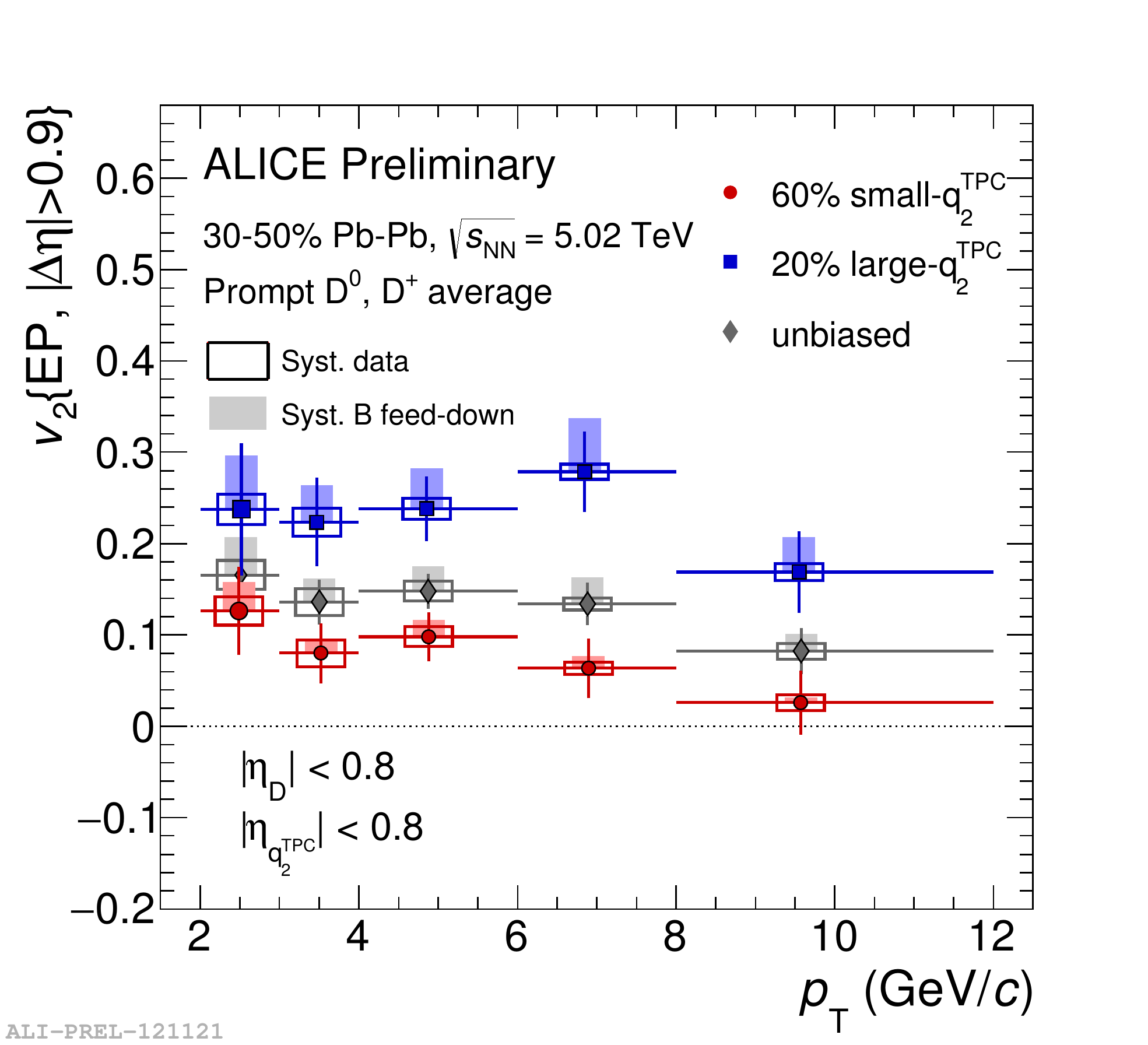}
\caption{Left: average prompt $\Dzero$, $\Dplus$, $\Dstar$ $v_2$ as a function of $\pt$ in Pb--Pb collisions at $\sqrtsNN=5.02~\TeV$ in the 30--50\% centrality class compared to the same measurement and the $\pi^{\pm}$ $v_2$ at $\sqrtsNN=2.76~\TeV$ \cite{Acharya:2017qps}. Right: average prompt $\Dzero$ and $\Dplus$ $v_2$ in 30--50\% Pb--Pb collisions at $\sqrtsNN=5.02~\TeV$ for the 60\% of events with smallest $q_2$ and the 20\% of events with largest $q_2$ compared to the unbiased result.}
\label{fig:Dmesonv2}       
\end{figure}
The left panel of Figure~\ref{fig:Dmesonv2} shows the prompt D-meson $v_2$ as a function of $\pt$ measured in Pb--Pb collisions at $\sqrtsNN=5.02~\TeV$ in the 30--50\% centrality class \cite{Acharya:2017qps}, using the Event-Plane method \cite{Poskanzer:1998yz}. The pseudorapidity gap ($|\Delta\eta|>0.9$) between the reconstructed D mesons and the particles used to estimate the Event-Plane angle suppresses the non-flow contributions in the $v_2$ arising from the correlation in azimuth of particles from decays or from jets. The $\Dzero$, $\Dplus$ and $\Dstar$ $v_2$ are consistent with each other and larger than zero in $2<\pt<10~\GeV/c$, indicating that low- and intermediate-$\pt$ charm quarks participate to the collective expansion of the medium. For $\pt>10~\GeV/c$ the D-meson $v_2$ is still positive, though comparable with zero. Furthermore, the D-meson $v_2$ at $\sqrtsNN=5.02~\TeV$ is found to be compatible to that measured at $\sqrtsNN=2.76~\TeV$ and similar in magnitude to that of charged pions in the same centrality class, as shown in the left panel of Figure~\ref{fig:Dmesonv2}. The elliptic flow coefficient of prompt $\Ds$ is also positive in $2<\pt<8~\GeV/c$ with a significance of about $2.6~\sigma$ and is compatible within uncertainties to that of non-strange D mesons \cite{Acharya:2017qps}. 
	The $\Dzero$ and $\Dplus$ $v_2$ were also investigated for the first time with an Event-Shape Engineering (ESE) technique. This method, consisting in measuring the D-meson $v_2$ for classes of events with different average elliptic flow, provides more insight into the coupling of the charm quark with the bulk of light quarks. The average ellipticity can be quantified by the magnitude of the second-harmonic reduced flow vector $q_2=|\pmb{Q}_2|/\sqrt{M}$ \cite{Voloshin:2008dg}, where $M$ is the multiplicity and $\pmb{Q}_2$ is the second-harmonic flow vector, whose components are given by $Q_{2,\rm x} = \sum_{i=1}^M \cos{2\varphi_i}$ and $Q_{2,\rm y} = \sum_{i=1}^M \sin{2\varphi_i}$. 
	The events in the 30--50\% centrality class were divided in the 60\% of events with smallest $q_2$ and the 20\% of the events with largest $q_2$. The average elliptic flow of prompt $\Dzero$ and $\Dplus$ mesons for the two $q_2$ classes of events compared to the unbiased one is presented in the right panel of Figure~\ref{fig:Dmesonv2}. The observation of a significant separation between the D-meson $v_2$ in the two classes of event shapes suggests that the charm quark is sensitive to the light-quark collectivity and to the event-by-event initial-state fluctuations. However, the effect could be slightly enlarged by autocorrelations between D mesons and $q_2$, since they are measured in the same pseudorapidity region $|\eta|<0.8$.
\section{Conclusions}
\label{sec:Conclusions}
The ALICE Collaboration measured the $\Raa$ and the $v_2$ of prompt $\Dzero$, $\Dplus$, $\Dstar$ and $\Ds$ mesons in Pb--Pb collisions at $\sqrtsNN=5.02~\TeV$. The D-meson $\Raa$ shows an increasing suppression from peripheral to central events. 
A hint of charm hadronisation via coalescence is provided by the observation of a higher $\Ds$ $\Raa$ with respect to that of non-strange D mesons. 
A significantly non-zero D-meson $v_2$ is measured for $2<\pt<10~\GeV/c$ in mid-central collisions, confirming the participation of the c-quark to the expanding dynamics of the hot medium observed at $\sqrtsNN=2.76~\TeV$. For the first time at the LHC, the $\Ds$ $v_2$ was measured and the ESE technique was applied to the non-strange D-meson elliptic flow.	

\bibliographystyle{utphys}
\bibliography{biblio}{}
\end{document}